\documentclass[copyright,creativecommons]{eptcs}
\providecommand{\event}{GandALF 2010} 
\usepackage{breakurl}             

\usepackage[latin1]{inputenc}
\usepackage{amsmath}
\usepackage{amssymb}
\usepackage{xspace}
\usepackage{color}
\usepackage{subfigure}
\usepackage{graphicx}
\usepackage{epic,eepic}
\usepackage{algorithm}
\usepackage{algpseudocode}

\newtheorem{theorem}{Theorem}[section]
\newtheorem{lemma}[theorem]{Lemma}

\newcommand{\ignore}[1]{}

\newcommand{\Nat}{\mathbb{N}}

\newcommand{\ran}{\mathtt{ran}}
\newcommand{\dom}{\mathtt{dom}}
\newcommand{\cons}[4]{\mathtt{Con}_{#1,#2}(#3,#4)}
\newcommand{\impr}[3]{\mathtt{Imp}_{#1,#2}(#3)}

\newcounter{linenumber}

\allowdisplaybreaks

\title{Local Strategy Improvement for Parity Game Solving}
\author{Oliver Friedmann 
\institute{Dept.\ of Computer Science \\ University of Munich, Germany}
\email{}
\and Martin Lange
\institute{Dept.\ of Computer Science \\ University of Kassel, Germany}
\email{}
}
\def\titlerunning{Local SI for Parity Game Solving}
\def\authorrunning{O.~Friedmann \& M.~Lange}

\begin{document}

\maketitle

\begin{abstract}
The problem of solving a parity game is at the core of many problems in model checking,
satisfiability checking and program synthesis. Some of the best algorithms for solving parity game
are strategy improvement algorithms. These are global in nature since they require the entire parity
game to be present at the beginning. This is a distinct disadvantage because in many applications one
only needs to know which winning region a particular node belongs to, and a witnessing winning
strategy may cover only a fractional part of the entire game graph.

We present a local strategy improvement algorithm which explores the game graph on-the-fly
whilst performing the improvement steps. We also compare it empirically with
existing global strategy improvement algorithms and the currently only other local algorithm
for solving parity games. It turns out that local strategy improvement can outperform these others
by several orders of magnitude.  
\end{abstract}

\section{Introduction}
\label{sec:intro}

Parity games are two-player games of perfect information played on labelled, directed graphs. They are
at the core of various problems in computer-aided verification, namely model checking
\cite{Stirling95}, satisfiability checking for branching-time logics \cite{FriedmannL:AutoTab09} and
controller synthesis \cite{AVW-TCS03}. They are also closely related to other games of infinite
duration, in particular mean and discounted payoff as well as stochastic games
\cite{Jurdzinski/98,Stirling95}.

A variety of algorithms for solving parity games has been invented so far. Among these, strategy 
improvement (SI) algorithms have proved to be very successful both because of their theoretical elegance
as well as their good performance in practice. The name hints at the general method according to which 
these algorithms work. Starting with an initial strategy, they compute an evaluation which either 
witnesses that the current strategy is winning, or it offers ways to improve the current strategy. SI 
is therefore in fact an algorithm scheme parametrised by a policy which determines how to improve
a non-winning strategy.
The first SI algorithm was given by Jurdzi{\'n}ski and V\"oge \cite{conf/cav/VogeJ00} which was
inspired by and improves over Puri's SI algorithms for parity and discounted payoff games
\cite{purithesis}. Later, Schewe presented another SI algorithm for parity games and payoff games
\cite{conf/csl/Schewe08}.

A disadvantage common to all these algorithms is their global nature: they require the entire game graph
to be present at the start of the algorithm's execution. They then compute for a given player his entire 
winning region and a corresponding winning strategy. This is a drawback because in many applications of
parity games taken from program verification one only needs to know for a specific node in the game which
winning region it belongs to, together with a winning strategy witnessing this. 

As an example, consider the model checking problem for branching-time temporal logics like CTL,
CTL$^*$, or the modal $\mu$-calculus on finite-state systems for instance. These can easily be regarded
as instances of the parity game solving problem \cite{Stirling95,LS-JLC2002} in which the nodes are
pairs of transition system states and (sets of) subformulas of the input formula. Solving the model
checking problem means solving the parity game locally, i.e.\ deciding which player wins the node
composed of the transition system's initial state and the input formula. A global algorithm would
compute this information for every (reachable) pair of nodes and (set of) subformulas. In model
checking, local solving is usually considered to be more desirable for explicitly represented state
spaces \cite{Bhat:1996:ELM}. Similarly, the satisfiability problem for branching-time logics reduces to 
the problem of solving parity games \cite{FriedmannL:AutoTab09} and, again, in order to decide whether or 
not the input formula is satisfiable one needs to know whether or not player $0$ wins a particular node of 
the resulting parity game. The winners of the other nodes are irrelevant.

It is known that the problem of solving a parity game is equivalent under linear-time reductions to
the model checking problem for the modal $\mu$-calculus. There are local model checkers, in particular
the one by Stevens and Stirling \cite{StevensStirling98} which can therefore be rephrased as a parity 
game solver. It is generally outperformed in practice by the global algorithms -- when the task is to
solve an entire game --  in particular the SI algorithms \cite{fl-atva09}. It is therefore fair to say 
that local parity game solvers with reasonable efficiency in practice are desirable but do not exist yet.

Here we present a local SI algorithm scheme for solving parity games. It solves a parity game locally,
i.e.\ starting with a given initial node it expands the game graph in an on-the-fly fashion and
intertwines this expansion with evaluations needed in order to improve current strategies. It terminates
as soon as it has determined which winning region the initial node belongs to. 

The rest of the paper is organised as follows. Sect.~\ref{sec:games} recalls parity games and global
SI. Sect.~\ref{sec:local} presents local SI. Sect.~\ref{sec:experiments} reports on empirical
comparisons between local SI, global SI, and the local model checking algorithm by Stevens/Stirling when
rephrased as a parity game solver. We show that local SI can outperform the others significantly. 

The contribution of this paper therefore lies in the practical use of parity game solvers. The presented
algorithm is not meant to improve other algorithms in asymptotic worst-case complexity. Its relevance is 
justified by the improvement in solving parity games in practice which this algorithm can bring along.


\section{Preliminaries}
\label{sec:games}

\subsection{Parity Games}
A \emph{parity game} is a tuple $G = (V,V_0,V_1,E,\Omega)$ where $(V,E)$ forms a directed graph
in which each node has at least one successor. The set of nodes is partitioned into
$V = V_0 \cup V_1$ with $V_0 \cap V_1 = \emptyset$, and $\Omega : V \to \Nat$ is the \emph{priority function}
that assigns to each node a natural number called the \emph{priority} of the node. We write
$|\Omega|$ for the index of the parity game, that is the number of different priorities assigned to
its nodes. The graph is not required to be total. Here we only consider games based on finite graphs.
W.l.o.g.\ we will assume that games are \emph{turn-based}, i.e.\ for
all $v \in V$ and all $i \in \{0,1\}$ we have: if $v \in V_i$ then $vE \subseteq V_{1-i}$. 

We also use infix notation $vEw$ instead of $(v,w) \in E$ and define the set of all \emph{successors} of
$v$ as $vE := \{ w \mid vEw \}$, as well as the set of all \emph{predecessors} of $w$ as
$Ew := \{ v \mid vEw \}$. 


The game is played between two players called $0$ and $1$ in the following way. Starting in a node
$v_0 \in V$ they construct a path through the graph as follows. If the construction so far
has yielded a finite sequence $v_0\ldots v_n$ and $v_n \in V_i$ then player $i$ selects a $w \in v_nE$
and the play continues with the sequence $v_0\ldots v_n w$. Otherwise, the play ends in a sink $v_n$.

Every infinite play has a unique winner given by the \emph{parity} of the greatest priority that occurs infinitely
often in a play. The winner of the play $v_0 v_1 v_2 \ldots$ is player $i$ iff
$\max \{ p \mid \forall j \in \Nat\ \exists k \geq j:\, \Omega(v_k) = p \} \equiv_2 i$ (where $i \equiv_2 j$
holds iff $|i - j| \mod 2 = 0$). That is, player $0$ tries to make an even priority occur infinitely often
without any greater odd priorities occurring infinitely often, player $1$ attempts the converse. For finite plays
$v_0 v_1 v_2 \ldots v_n$ that end in a sink, the winner is given by the \emph{chooser} of $v_n$. Player $i$
wins the play iff $v_n \in V_{1-i}$.

A \emph{positional strategy} for player $i$ in $G$ is a --- possibly partial --- function $\sigma: V_i \to V$.
A play $v_0 v_1 \ldots$ \emph{conforms} to a strategy $\sigma$ for player $i$ if for all $j \in \Nat$ we
have: if $v_j \in V_i$ then $v_{j+1} = \sigma(v_j)$. Intuitively, conforming to a strategy means to always
make those choices that are prescribed by the strategy. A strategy $\sigma$ for player $i$ is a
\emph{winning strategy} in node $v$ if player $i$ wins every play that begins in $v$ and conforms to $\sigma$.
We say that player $i$ \emph{wins} the game $G$ starting in $v$ iff he/she has a winning strategy for $G$
starting in $v$. Let $\bot$ denote the empty strategy.

With $G$ we associate two sets $W_0,W_1 \subseteq V$; $W_i$ is the set of all nodes $v$ s.t.\ player
$i$ wins the game $G$ starting in $v$. Here we restrict ourselves to positional strategies because it is
well-known that a player has a (general) winning strategy iff she has a positional winning strategy for a
given game. In fact, parity games enjoy positional determinacy meaning that for every node
$v$ in the game either $v \in W_0$ or $v \in W_1$ \cite{focs91*368}. Furthermore,
it is not difficult to show that, whenever player $i$ has winning strategies $\sigma_v$ for all $v \in U$ for
some $U \subseteq V$, then there is also a single strategy $\sigma$ that is winning for player $i$ from every
node in $U$.

The problem of \emph{(globally) solving a parity game} is to decide for every node $v \in V$ whether ot not it 
belongs to $W_0$. 
%
The problem of \emph{locally solving a parity game} is to decide, for a given node $v$, whether or not 
$v$ belongs to $W_0$. Note that it may not be necessary to visit all nodes of a game in order to answer
this question. Consider the following trivial example. The two nodes under consideration are connected
to each other and have no further outgoing edges. Hence, this simple cycle is clearly won by one of the
two players. A depth-first search can find this loop whilst leaving the rest of the game unexplored.


A strategy $\sigma$ for player $i$ induces a \emph{strategy subgame} $G|_\sigma := (V, V_0, V_1, E|_\sigma, \Omega)$
where $E|_\sigma := \{(u, v) \in E \mid u \in V_i \Rightarrow (u \in dom(\sigma) \wedge \sigma(u) = v)\}$. The set
of strategies for player $i$ is denoted by $\mathcal{S}_i(G)$.

Let $U \subseteq V$, $i \in \{0,1\}$ and $\sigma$ be an $i$-winning strategy on $U$. The $i$-attractor of $U$
is the least set $W$ s.t.\ $U \subseteq W$ and whenever $v \in V_i$ and $vE \cap W \ne \emptyset$, or $v \in V_{1-i}$
and $vE \subseteq W$ then $v \in W$. Hence, the $i$-attractor of $U$ contains all nodes from which player $i$ can move
``towards'' $U$ and player $1-i$ must move ``towards'' $U$. At the same time, it is possible to construct an
\emph{attractor strategy} which is a positional strategy in a
reachability game. Following this strategy guarantees player $i$ to reach a node in $U$ eventually, regardless
of the opponent's choices. Let $\textsc{Attr}_{G,i}(U,\sigma) = (W,\tau)$ denote the $i$-attractor of $U$ s.t.\
$\tau$ is an $i$-winning strategy on $W$. Note that the $i$-attractor can be computed in
$\mathcal{O}(|W \cup \bigcup_{v \in W} Ev|)$.

An important property that has been noted before \cite{TCS::Zielonka1998,Stirling95} is that removing
the $i$-attractor of any set of nodes from a game will still result in a total game graph.


\subsection{Global Strategy Improvement}
\label{sec:global}

We recap global SI \cite{conf/cav/VogeJ00,conf/csl/Schewe08}. It defines a valuation of strategies and a 
partial order on them. The algorithm then starts with an arbitrary strategy, computes its valuation, 
improves it accordingly and iterates this until no improvement is possible anymore, which means that
winning regions and strategies have been found.

Strategy valuations can be defined in different ways, but usually through \emph{node valuations} for
each node $v$. The latter are totally ordered and induced by a path starting in $v$ leading to a cycle.
The valuation process then selects for each node $v$ the node valuation associated with a path $\pi$
ending in a cycle starting in $v$ conforming to the strategy $\sigma$ that yields the worst valuation.


Schewe \cite{conf/csl/Schewe08} uses a simple valuation by considering an \emph{escape sink}
transformation of the given parity game: all priorities are increased by $2$, two new self-cycle nodes
with priorities $0$ and $1$ are added to the graph and each player 0 node is connected to the $1$ and each
player 1 node is connected to the $0$. The transformed game is equivalent to the original game, 
but now the only occurring cycle nodes in valuations are 
$0$ and $1$ (assuming that one starts with a reasonable initial strategy). 
Schewe's node valuation therefore is either the set of nodes leading to the unprofitable self-cycle or
$\infty$, indicating that the node is already won by the $\sigma$-player following $\sigma$.

We follow this approach, but instead of adding additional self-cycles, we handle partial
strategies $\sigma$: whenever the strategy is undefined for a certain node $v$, we identify
this with the decision to move to the imaginary self-cycle.


Let $G$ be a parity game. The \emph{relevance ordering} $<$ on $V$ is induced by $\Omega$:
$v < u :\iff \Omega(v) < \Omega(u)$; we extend $<$ w.l.o.g.\ to an arbitrary but fixed
total ordering. Additionally one defines the total \emph{reward ordering w.r.t.\ player $q$} $\prec_q$
on $V$: $v \prec_q u$ holds iff one of the following conditions applies.
\begin{enumerate}
\item either $\Omega(u) \equiv_2 q$ and $v < u$
\item or $\Omega(u) \equiv_2 q$ and $\Omega(v) \not\equiv_2 q$
\item or $\Omega(u) \not\equiv_2 q$, $\Omega(v) \not\equiv_2 q$ and $u < v$
\end{enumerate}
%
The length of a loopless path $\pi$ is denoted by $|\pi|$. The set of loopless paths $\pi$ in 
$G$ starting in $v$ is denoted by $\Pi_G(v)$. 

The set of \emph{node valuations w.r.t.\ $G$}, $\mathcal{V}_G$,
is defined as $\mathcal{V}_G := \{\infty\} \cup 2^V$, i.e.\ a node valuation is either $\infty$
or a set of nodes. Given a player $q$, we extend the reward ordering to a total ordering on node
valuations. Let $M, N \subseteq V$ with $M \not= N$.
\begin{displaymath}
M \prec_q N  \, : \iff \,
\begin{cases}
\left(max_<(M \triangle N) \in N \textrm{ and } max_<(M \triangle N) \equiv_2 q \right) \textrm{ or } \\
\left(max_<(M \triangle N) \in M \textrm{ and } max_<(M \triangle N) \not\equiv_2 q \right)
\end{cases}
\end{displaymath}
where $M \triangle N$ denotes the symmetric difference of both sets. Additionally $M \prec_q \infty$
holds for every $M \subseteq V$.

A loopless non-empty path $\pi = v_0\ldots v_{k-1}$ induces a \emph{$q$-node valuation for the node $v_0$}
as follows.
\begin{displaymath}
\vartheta_q(\pi) := \begin{cases}
\infty & \textrm{if } v_{k-1} \in V_{1-q} \\
\{v_i \mid 0 \leq i < k\} & \textrm{otherwise}
\end{cases}
\end{displaymath}
Let $\vartheta$ be a node valuation and $v$ be a node. We define an addition operation $\vartheta \oplus v$
as follows.
\begin{displaymath}
\vartheta \oplus v := \begin{cases}
\infty & \textrm{if } \vartheta = \infty \textrm{ or } v \in \vartheta \\
\vartheta \cup \{v\} & \textrm{otherwise}
\end{cases}
\end{displaymath}
A \emph{game valuation} is a map $\Xi: V \rightarrow \mathcal{V}_G$ assigning
to each $v \in V$ a node valuation. A partial ordering on game valuations is defined as follows:
\begin{displaymath}
\Xi \lhd_q \Xi' \, : \iff \, \left(\Xi(v) \preceq_q \Xi'(v) \textrm{ for all } v \in V\right) \textrm{ and } \left(\Xi \not= \Xi'\right)
\end{displaymath}
A strategy $\sigma$ of player $q$ therefore can be evaluated as follows:
\begin{displaymath}
\Xi_{q,\sigma}:\, v \mapsto \min_{\prec_q} \{\vartheta_q(\pi) \mid \pi \in \Pi_{G|_\sigma}(v)\}
\end{displaymath}
%
%
A game valuation $\Xi$ induces a total counter-strategy $\tau_\Xi$ of player $1-q$ by selecting
the least profitable strategy decision with respect to $\Xi$:
\begin{displaymath}
\tau_\Xi: v \in V_{1-q} \mapsto \min_{\prec_q} U_\Xi(v)
\end{displaymath}
where $U_\Xi(v) = \{u \in vE \mid \forall w \in vE: \Xi(u) \preceq_q \Xi(w)\}$.

If $\Xi$ originates from a strategy $\sigma$, $\tau_\Xi$ can be seen as the best counter-strategy
against $\sigma$; we also write $\tau_\sigma$ for $\tau_{\Xi_{\sigma}}$.

Given a game valuation $\Xi$ and a strategy $\sigma$ for player $q$, the \emph{set of $\Xi$-consistent nodes},
$\cons{G}{q}{\sigma}{\Xi}$ is defined as follows.
\begin{flalign*}
\cons{G}{q}{\sigma}{\Xi} \enspace := \enspace &\{v \in V_0 \cap \dom(\sigma) \mid \Xi(v) = \Xi(\sigma(v)) \oplus v\} \cup \\
                            &\{v \in V_0 \setminus \dom(\sigma) \mid \Xi(v) = \{v\}\} \cup \\
							&\{v \in V_1 \mid \Xi(v) = \min_{\prec_q} \{\Xi(u) \oplus v \mid u \in vE\}
\end{flalign*}
Note that for every strategy $\sigma$ it holds that $\cons{G}{q}{\sigma}{\Xi_\sigma} = V$.

A valuation $\Xi$ originating from a strategy $\sigma$ can be used to create a new strategy of player $q$.
The strategy improvement algorithm only allows to select new strategy decisions for player $q$ occurring in
the \emph{improvement arena} $\mathcal{A}_{G,q,\sigma} := (V,\ V_0,\ V_1,\ E',\ \Omega)$ where
\begin{displaymath}
vE'u \, : \iff \,
vEu \textrm{ and } \left(v \in V_{1-q} \textrm{ or } (v \in V_q \textrm{ and } \Xi(\sigma(v)) \preceq_q \Xi(u))\right)
\end{displaymath}
Thus all edges performing worse than the current strategy are removed from the game. A player $q$ strategy $\sigma$
is \emph{$q$-improvable} iff there is a node $v \in V_q$, a node $u \in V$ with $vEu$ and $\sigma(v) \not= u$ s.t.\
$\Xi(\sigma(v)) \prec_q \Xi(u)$. The \emph{set of improvable nodes} w.r.t.\ a valuation $\Xi$ is defined as follows.
\begin{displaymath}
\impr{G}{q}{\Xi} := \{v \in V_0 \mid \Xi(v) < \max_{\prec_q} (\{\{v\}\} \cup \{\Xi(u) \oplus v \mid u \in vE\})
\end{displaymath}
An \emph{improvement policy} now selects a strategy for player $q$ in a given improvement arena w.r.t.\
a valuation originating from a strategy. More formally: an improvement policy is a map
$\mathcal{I}_{G,q}: \mathcal{S}_q(G) \rightarrow \mathcal{S}_q(G)$ fulfilling the following two conditions
for every player $q$ strategy $\sigma$.
\begin{enumerate}
\item For every node $v \in V_q$ it holds that $(v,\mathcal{I}_{G,q}(\sigma)(v))$ is an edge in $\mathcal{A}_{G,q,\sigma}$.
\item If $\sigma$ is $q$-improvable then there is a node $v \in V_q$ s.t.\ $\Xi_\sigma(\sigma(v)) \prec_q \Xi_\sigma(\mathcal{I}_G(\sigma)(v))$.
\end{enumerate}
Updating a strategy according to an improvement policy can only result in strategies with better valuations 
(w.r.t.\ $\lhd_q$) \cite{conf/cav/VogeJ00}.
%
%
If a strategy $\sigma$ is not $q$-improvable, the strategy iteration comes to an end. Winning regions and
strategies have then been found in the following sense. 
%
%
$W_q = \{v \mid \Xi_{G,q,\sigma}(v) = \infty\}$; 
$W_{1-q} = \{v \mid \Xi_{G,q,\sigma}(v) \not= \infty\}$; 
$\sigma$ is a winning strategy for player $q$ on $W_q$;
$\tau := \tau_\sigma$ is a winning strategy for player $1-q$ on $W_{1-q}$.

SI w.r.t.\ player $q$ starts with the empty initial strategy $\iota_G$ and
updates it according to a given improvement policy $\mathcal{I}_G$ until it becomes non-improvabable.

\begin{algorithm}
\caption{Strategy Iteration}
\begin{algorithmic}[1]
\State $\sigma \gets \iota_G$
\While {$\sigma$ is $q$-improvable}
	\State $\sigma \gets \mathcal{I}_G(\sigma)$
\EndWhile
\State \textbf{return} $W_q$, $W_{1-q}$, $\sigma$ and some $\tau$ as described above
\end{algorithmic}
\end{algorithm}


\section{A Local Strategy Improvement Algorithm}
\label{sec:local}

Note that global SI is formulated w.r.t.\ one player $q$: starting with a strategy for player $q$
it iteratively improves it until it becomes a winning strategy on some part of the game graph.
A winning strategy for the opponent on the complementary part is then derived from this. This is 
counterproductive for local solving: if local SI was done for one player only, and the opponent is 
the winner of the designated input node, then the algorithm would have to expand the entire 
reachable part of the game graph and it would therefore be at most as good as global SI. 
Consequently, local SI is parametrised by an improvement policy.

We fix a parity game $G = (V,V_0,V_1,E,\Omega)$ and an initial position
$v^* \in V$ for which the winner and an associated winning strategy is to be computed. The algorithm
maintains two sets $W_0$ and $W_1$ in which it collects the winning nodes for players $0$ and $1$,
resp., as well as two winning strategies $\varrho_0$ and $\varrho_1$ on these sets. Initially, all
of them are empty.

Local SI is carried out for both players alternatingly on potentially different subgraphs of $G$.
Although it would be possible to run the strategy iteration only
for one player, say 0, it would require the algorithm to expand the complete graph that is
reachable from $v^*$ if player 1 wins the initial position $v^*$. Therefore, the algorithm
basically alternates the strategy iteration w.r.t.\ both players indepedently.

For each player $q$, it maintains initially empty node sets $U_q$ --- called the \emph{$q$-subgraph}
--- and $E_q$ initialised to $\{v^*\}$ --- called the \emph{$q$-expansion set} --- with the following 
properties. The $q$-subgraph represents a local view onto the game $G$ for player $q$. The 
$q$-expansion set contains nodes which lie just outside of the $q$-subgraph and which can be used
in order to expand the latter. These components satisfy the following invariants.
\begin{enumerate}
\item \emph{Subgraph-Closure} (SC): for every $v \in U_q \cap V_{1-q}$ it holds that $vE \setminus (W_0
  \cup W_1) \subseteq U_q$; this ensures that for every opponent node in the $q$-subgraph all
  possible choices are also included in the set.

\item \emph{Winning-Closure} (WC): $\textsc{Attr}_{G|_{U_0 \cup U_1},q}(W_q,\varrho_q) =
  (W_q,\varrho_q)$; this ensures that the set of known winning nodes is closed under attractor
  strategies w.r.t.\ the known subgraphs.

\item \emph{Winning-Exclusion} (WE): $U_q \cap (W_0 \cup W_1) = \emptyset$; this ensures that
  both winning sets cannot be reached by a trivial strategy by one of the two players.

\item \emph{Border-Expansion} (BE): $\{u \in vE \setminus (W_0 \cup W_1 \cup U_q)\mid v \in U_q\}
  \subseteq E_q \subseteq V \setminus (W_0 \cup W_1 \cup U_q)$; this ensures that all nodes
  which are not in the $q$-subgraph but can be reached from it via one edge are in $E_q$.
\end{enumerate}
Additionally, for each player $q$ we maintain a set $I_q$ --- called the \emph{$q$-improvement set}
---, a valuation $\Xi_q$ for $G|_{U_q}$, a $q$-strategy $\sigma_q$
and another set $C_q$ --- called the \emph{$q$-change set}. They satisfy the following invariants.
\begin{enumerate}
\item \emph{Subgame-Strategy} (SS): $\dom(\sigma_q) \cup \ran(\sigma_q) \subseteq U_q$; this ensures
  that the strategy associated with the subgame is not leading out of the subgame itself.

\item \emph{Valuation-Consistency} (VC): $U_q \setminus \cons{G}{q}{\sigma_q}{\Xi_q} \subseteq C_q
  \subseteq U_q$; this ensures that the $q$-change set contains at least all these nodes of the
  $q$-subgraph that have an inconsistent valuation.

\item \emph{Improvement-Covering} (IC): $I_q \setminus C_q = \impr{G}{q}{\Xi_q} \setminus C_q$; this
  ensures that the $q$-improvement set contains all improvement nodes of the $q$-subgraph
  disregarding the $q$-change set.
\end{enumerate}
Initially, $I_q$, $C_q$ as well as $\Xi_q$ and $\sigma_q$ are empty. Finally, the algorithm maintains
a \emph{current player} $q^*$ initialised arbitrarily.

It is parametrized by the following policy that is allowed to access all data that
is maintained by the algorithm.
\begin{enumerate}


\item \emph{Improvement Policy}: a function $\textsc{ImprPol}$ that returns a non-empty
  set of pairs $(v,u)$ s.t.\ $v \in U_{q^*} \cap V_{q^*}$, $u \in U_{q^*}$ and $(v,u)$ is a proper
  improvement edge w.r.t.\ $\Xi_{q^*}$.
\end{enumerate}
Whenever the algorithm encounters a new set of winning nodes, it calls the function 
$\textsc{Winning}$ with parameters $q$, $W$ and $\varrho$ where $q$ is the winning player, $W$ is the 
new set of nodes won by $q$ and $\varrho$ is an associated winning strategy on $W$.
It computes the $q$-attractor of the given winning set in the subgraph $U_0 \cup U_1$, adds it to
the winning set of $q$ and updates all other sets accordingly. In particular, the change sets
are updated to contain the border nodes in the respective subgraph.

\renewcommand{\textfraction}{0.1}
\renewcommand{\topfraction}{0.99}
\setcounter{topnumber}{3}
\setcounter{totalnumber}{3}

In order to refer to the data before and after the call, we refer to the unchanged name of a data variable 
for its value before the call and to the $\dagger$-superscript for its value after the call.

\begin{lemma}
Let (SC), (WC), (WE), (BE), (VC) and (IC) hold for both players. Let $q$ be a player, $W \subseteq U_q$ be a set and $\varrho$
be a winning strategy for $q$ on $W$. Then, calling $\textsc{Winning}(q, W, \varrho)$ preserves (SC), (WC), (WE),
(BE), (VC) and (IC) for both players. Additionally,
$C^\dagger_{q'} = (C_{q'} \setminus W^\dagger_q) \cup (U_{q'} \cap \bigcup_{v \in W^\dagger_q \setminus W_q} Ev)$
for both players $q'$ and $W \subseteq W^\dagger_q$.

Also, $W_{q'} \subseteq W^\dagger_{q'}$ and $U_{q'} \subseteq U^\dagger_{q'} \cup W^\dagger_{q'}$ for both players $q'$.
\end{lemma}

\begin{algorithm}[t]
\caption{Update of Winning Set}
\begin{algorithmic}[1]
\Procedure{Winning}{$q$, $W$, $\varrho$}
	\State $(W', \varrho') \gets $\textsc{Attr}$_{G|_{U_0 \cup U_1},q}(W,\varrho)$
	\State $W_q \gets W_q \cup W'$
	\State $\varrho_q \gets \varrho_q \cup \varrho'$
	\State $B \gets \bigcup_{v \in W'} Ev$
	\State $(I_0, I_1) \gets (I_0 \setminus W', I_1 \setminus W')$
	\State $(E_0, E_1) \gets (E_0 \setminus W', E_1 \setminus W')$
	\State $(C_0, C_1) \gets ((C_0 \setminus W') \cup (U_0 \cap B), (C_1 \setminus W') \cup (U_1 \cap B))$
	\State $(U_0, U_1) \gets (U_0 \setminus W', U_1 \setminus W')$
	\State $(\Xi_0, \Xi_1) \gets (\Xi_0 \setminus W', \Xi_1 \setminus W')$
	\State $(\sigma_0, \sigma_1) \gets (\sigma_0 \setminus W', \sigma_1 \setminus W')$
\EndProcedure
\end{algorithmic}
\end{algorithm}

\begin{algorithm}[t]
\caption{Expansion of Nodes}
\begin{algorithmic}[1]
\Procedure{Expand}{$q$, $N$}
	\State $(W'_0, \varrho'_0) \gets (\emptyset, \bot)$
	\State $(W'_1, \varrho'_1) \gets (\emptyset, \bot)$
	\While {$N$ is not empty}
		\State $v \gets $ pick from $N$
		\State $N \gets N \setminus \{v\}$
		\State $U_q \gets U_q \cup \{v\}$
		\State $\Xi_q(v) \gets \emptyset$
		\State $E_q \gets E_q \setminus \{v\}$
		\State $C_q \gets C_q \cup \{v\}$
		\State $c \gets $ chooser of $v$
		\If {there is some $u \in W_c \cup W'_c$ s.t.\ $vEu$}
			\State $W'_c \gets W'_c \cup \{v\}$
			\State $\varrho'_c \gets \varrho'_c[v \mapsto u]$
		\ElsIf {$u \in W_{1-c} \cup W'_{1-c}$ for every $u$ with $vEu$}
			\State $W'_{1-c} \gets W'_{1-c} \cup \{v\}$
		\ElsIf {$c = q$}
			\State $E_q \gets E_q \cup (vE \setminus (U_q \cup W_0 \cup W'_0 \cup W_1 \cup W'_1))$
		\Else
			\State $N \gets N \cup (vE \setminus (U_q \cup W_0 \cup W'_0 \cup W_1 \cup W'_1))$
		\EndIf		
	\EndWhile
	\State \textsc{Winning}($0$, $W'_0$, $\varrho'_0$)
	\State \textsc{Winning}($1$, $W'_1$, $\varrho'_1$)
\EndProcedure
\end{algorithmic}
\end{algorithm}

\begin{algorithm}[!t]
\caption{Local Evaluation}
\begin{algorithmic}[1]
\Procedure{Evaluate}{$q$}
	\State $\tau \gets \bot$
	\State $W \gets \emptyset$
	\State $D \gets \emptyset$
	\While {$C_q$ is not empty}
		\State $v \gets $ pick from $C_q$
		\State $C_q \gets C_q \setminus \{v\}$
		\State $\vartheta \gets \{v\}$
		\If {$v \in V_q \textbf{ and } v \in \dom(\sigma_q)$}
			\State $\vartheta \gets \Xi_q(\sigma_q(v)) \oplus v$
		\ElsIf {$v \in V_{1-q}$}
			\State $\vartheta \gets \min_{\prec_q} \{\Xi_q(u) \oplus v \mid u \in vE \cap U_q\}$
		\EndIf
		\If {$\vartheta \not= \Xi_q(v)$}
			\State $\Xi_q(v) \gets \vartheta$
			\State $D \gets D \cup \{v\}$
			\State $C_q \gets C_q \cup (Ev \cap U_q)$
		\EndIf
	\EndWhile
	\State $D \gets D \cup \bigcup_{v \in D} Ev$
	\While {$D$ is not empty}
		\State $v \gets \textrm{pick from } D$
		\State $D \gets D \setminus \{v\}$
		\If {$\Xi_q(v) = \infty$}
			\State $I_q \gets I_q \setminus \{v\}$
			\State $W \gets W \cup \{v\}$
			\If {$v \in V_q$}
				\State $\tau \gets \tau[v \mapsto \sigma_q(v)]$
			\EndIf
		\ElsIf {$v \in V_q$}
			\If {$\textrm{there is some } u \in vE \cap U_q \textrm{ with } \Xi_q(\sigma_q(v)) \prec_q \Xi_q(u)$}
				\State $I_q \gets I_q \cup \{v\}$
			\Else
				\State $I_q \gets I_q \setminus \{v\}$
			\EndIf
		\EndIf
	\EndWhile
	\State \textbf{return} $W$ and $\varrho$
\EndProcedure
\end{algorithmic}
\end{algorithm}

Local SI expands the subgraph $U_{q^*}$ of the current player $q^*$ if there are no more improvement
edges in $U_{q^*}$ and the winner of $v^*$ has not been decided yet. Then, it calls a function
$\textsc{Expand}$ with parameters $q$ and $N$, where $q$ is a player with a non-empty expansion set and
$N$ is a non-empty subset of the expansion set.  This essentially adds the minimal set $N'$ to $U_q$
that contains $N$ s.t.\ the new $U_q$ fulfills the Subgraph-Closure property again. Additionally, the
$q$-change set is increased by the set of new nodes; it may happen, that expanding $U_q$ reveals new
edges that lead to nodes in the winning sets $W_0$ or $W_1$. The $\textsc{Expand}$ routine puts these
nodes aside and calls the $\textsc{Winning}$ procedure in the end to update the winning sets.
Therefore, it is possible that calling $\textsc{Expand}$ actually decreases the size of $U_q$.

\begin{lemma}
Let (SC), (WC), (WE), (BE), (VC) and (IC) hold for both players. Let $q$ be a player and $N \subseteq E_q$ be a set.
Then, calling $\textsc{Expand}(q, N)$ preserves (SC), (WC), (WE), (BE), (VC) and (IC) for both players. Additionally,
$N \subseteq U^\dagger_q \cup W^\dagger_0 \cup W^\dagger_1$ and
$C^\dagger_{q'} = (C_{q'} \setminus W^\dagger_q) \cup (U_{q'} \cap \bigcup_{v \in W^\dagger_q \setminus W_q} Ev) \cup (U^\dagger_{q'} \setminus U_{q'})$
for both players $q'$.

Also, $W_{q'} \subseteq W^\dagger_{q'}$ and $U_{q'} \subseteq U^\dagger_{q'} \cup W^\dagger_{q'}$ for both players $q'$.
\end{lemma}

Whenever the change set $C_q$ of a player $q$ is changed, a re-evaluation process needs to take place
that updates the valuation $\Xi_q$ accordingly. The local evaluation for player $q$ can be invoked by
calling \textsc{Evaluate}. During this process, it is possible
that some nodes of $U_q$ are identified as being won by player $q$. The function therefore returns a
set of nodes newly won by player $q$ as well as a winning strategy on this set.
It updates all dependent node valuations as long as their valuation is not 
locally consistent. The improvement set is then updated accordingly and the set of nodes that are won 
by the current valuation is returned along with the corresponding strategy.

\begin{lemma}
Let (SC), (WC), (WE), (BE), (VC) and (IC) hold for both players and let $q$ be a player.
Then, calling $\textsc{Evaluate}(q)$ terminates and -- returning $W$ and $\varrho$ -- preserves (SC), (WC), (WE), (BE), (VC)
and (IC) for both players. Additionally, $\varrho$ is a winning strategy of $q$ on $W$ and $C_q = \emptyset$.

Also, $W_{q'} \subseteq W^\dagger_{q'}$ and $U_{q'} \subseteq U^\dagger_{q'} \cup W^\dagger_{q'}$ for both players $q'$.
\end{lemma}

\begin{algorithm}[!t]
\caption{Main}
\begin{algorithmic}[1]
	\While {$v^* \not\in W_0 \cup W_1$}
		\If {$E_{q^*} = \emptyset \textbf{ and } I_{q^*} = \emptyset$}
			\State $W_{1-{q^*}} \gets W_{1-{q^*}} \cup U_{q^*}$
			\State $\varrho_{1-{q^*}} \gets \varrho_{1-{q^*}} \cup \tau_{G|_{U_{q^*}},\sigma_{q^*}}$
		\Else
			\If {$I_{q^*} = \emptyset$}
				\State $N \gets $pick non-empty subset of $E_{q^*}$
				\State \textsc{Expand}$({q^*},N)$
			\Else
				\State $I \gets $\textsc{ImprPol}$(\mathrm{data})$
				\ForAll {$(v,u) \in I$}
					\State $\sigma_{q^*}(v) \gets u$
					\State $C_{q^*} \gets C_{q^*} \cup \{v\}$
					\State $q^* \gets 1-q^*$
				\EndFor
			\EndIf
			\Repeat
				\State $\mathrm{stable} \gets \mathbf{true}$
				\For {$i = 0,1$}
					\State $(W,\varrho) \gets $\textsc{Evaluate}$(i)$
					\If {$W \not= \emptyset$}
						\State $\mathrm{stable} \gets \mathbf{false}$
						\State \textsc{Winning}$(i,W,\varrho)$
					\EndIf
				\EndFor
			\Until {$\mathrm{stable}$}
		\EndIf
	\EndWhile
\end{algorithmic}
\end{algorithm}

Local SI is then realised by the procedure \textsc{Main} which runs for as long as the
initial node $v^*$ is not contained in one of the two winning sets. If both the improvement set and the
expansion set of the current player are empty we know that the other player wins the whole subgraph.
Then it adds the entire subgraph to the opponent's winning set.
Otherwise, either one of the two sets is non-empty. If the improvement set is empty, it expands the
subgraph according to the picked non-empty set. Otherwise it improves the subgraph according to the
improvement policy. Additionally, switching between the two players is consulted.
Now, it may be the case that the valuation for one of the two players needs to be updated. Since an 
update of the valuation of one player may affect the winning sets of both and hence may require another 
update of either player's valuation afterwards, it re-evaluates for as long as the two change sets are 
non-empty.

\begin{lemma}
Calling $\textsc{Main}$ terminates s.t.\ there is one player $q$ with $v^* \in W_q$ and $\varrho_q$ being a
winning strategy for $q$ on $W_q$.
\end{lemma}


\section{Experimental Results}
\label{sec:experiments}

We report on the performance of the local SI algorithm in practice. We compare it in particular
to two other algorithms for parity games: global SI by Jurdzi{\'n}ski and V\"oge \cite{conf/cav/VogeJ00} 
and Stevens and Stirling's local $\mu$-calculus model checker \cite{StevensStirling98}. 
Local SI is supposed to remedy one of global SI's distinct disadvantages, hence this comparison. 
Note that both global SI algorithms \cite{conf/cav/VogeJ00,conf/csl/Schewe08} perform quite well in 
practice although Zielonka's recursive algorithm \cite{TCS::Zielonka1998} is usually even a bit better 
\cite{fl-atva09}. However, it is the case
that, whenever local SI is faster than global SI, it is also faster than the recursive algorithm.
We also remark that the recursive algorithm is inherently global in
nature, and we know of no idea how to create a local variant of this one.
Furthermore, the parity game solving algorithm derived from the local model checking algorithm is
of course a local algorithm already. It is therefore also natural contender for local SI.  
Previous tests have shown that it can be much slower than the
global algorithms \cite{fl-atva09}. This comparison therefore shows that this is not due
to the locality but to the solving technique. 

Local SI has been implemented in \textsc{PGSolver},\footnote{\small \tt
  http://www.tcs.ifi.lmu.de/pgsolver} a publicly available tool for solving parity games which is
written in OCaml and which contains implementations of the two competing algorithms as well as many
others. Furthermore, \textsc{PGSolver} uses a specialised routine for global solving called 
\emph{generic solver}. It does not feed the entire game graph to the chosen solving algorithm but
performs optimising pre-processing steps like decomposition into SCCs and removal of trivial 
winning regions etc. These greatly speed up global solving \cite{fl-atva09}. The runtime results
of the global algorithm reported here are obtained with all optimising pre-processing steps. On
the other hand, these steps do not make sense in the context of local solving because the removal
of trivial winning regions for instance is in fact a global operation. Hence, we do in fact compare
local SI as it is against global SI embedded in this optimisaing framework.

Local SI is run in these experiments with the standard all-max policy on the respective
improvement sets.

All tests have been carried out on a 64-bit machine with four quad-core Opteron\texttrademark{}
CPUs and 128GB RAM space. As benchmarks we use some hand-crafted verification problems and random
games. We report on the times needed to solve the games
(globally vs.\ locally) as well as the number of nodes in the games that have been visited. This
is an important parameter indicating the memory consumption of the algorithms on these benchmarks.
The times for global SI are composed of the time it takes to generate the parity game (which
is not part of the solving, strictly speaking, but needs to be considered too in comparison to
the local algorithms since they generate the game while they are solving it) and the time it takes 
to perform the strategy iteration.  
We only present instances of non-negligible running times. On the other hand, the solving of larger
instances not presented in the figures anymore has experienced time-outs after 
one hour, marked $\dagger$.

\begin{figure}[h]
\centering
{\scriptsize
\begin{tabular}[ht]{|l|r|r||r|r|r||r|r||r|r|}
 \hline
 \multicolumn{3}{|c||}{} & \multicolumn{3}{c||}{\rule[-2mm]{0pt}{6mm} global SI} & \multicolumn{2}{c||}{ local MC} & \multicolumn{2}{c|}{ local SI} \\
  & $n$ & \multicolumn{1}{c||}{$|\mathrm{Game}|$} & \multicolumn{1}{c|}{$t_\mathrm{generate}$} & 
  \multicolumn{1}{c|}{$t_\mathrm{iterate}$} & \multicolumn{1}{c||}{$t_\mathrm{solve}$} & \multicolumn{1}{c|}{visited} & \multicolumn{1}{c||}{$t_\mathrm{solve}$} & \multicolumn{1}{c|}{visited} & \multicolumn{1}{c|}{$t_\mathrm{solve}$} \\
 \hline\hline
 & $3$ & $603$ & $0.02$s & $0.00$s & $0.02$s & $603$ & $0.12$s & $603$ & $0.28$s \\
 & $4$ & $3,120$ & $0.08$s & $0.02$s & $0.10$s & $3,120$ & $2.32$s & $3,120$ & $8.66$s \\
 \enspace FIFO & $5$ & $19,263$ & $0.66$s & $0.25$s & $0.91$s & $19,263$ & $206.32$s & $19,263$ & \ $414.55$s \\
 \enspace elevator \enspace & $6$ & $138,308$ & $5.64$s & $2.68$s & $8.32$s & $-$ & $\dagger$ & $-$ & $\dagger$ \\
 & $7$ & $1,130,884$ & $57.57$s & $35.94$s & $93.51$s & $-$ & $\dagger$ & $-$ & $\dagger$ \\
\hline\hline
 & $5$ & $20,126$ & $0.80$s & $0.34$s & $1.14$s & $1,237$ & $0.02$s & $404$ & $0.03$s \\
 \enspace LIFO & $6$ & $142,720$ & $7.30$s & $4.13$s & $11.43$s & $3,195$ & $0.07$s & $581$ & $0.06$s \\
 \enspace elevator & $7$ & $1,154,799$ & $83.45$s & $66.91$s & $150.36$s & $22,503$ & $0.66$s & $806$ & $0.09$s \\
 & $8$ & \ $10,505,651$ & $2342.61$s & $3596.42$s & $5939.03$s  & $60,247$ & $2.11$s & $1,085$ & $0.15$s \\
  \hline
\end{tabular}}
\caption{$\mu$-Calculus Model Checking}
\label{fig:mucalculusmodelchecking}
\end{figure}

\paragraph*{$\mu$-Calculus Model Checking.}
As first benchmark we use the problem of verifying an elevator system. It serves $n$ floors and
maintains a list of requests which may contain each floor at most once. A detailed
description can be found in \cite{fl-atva09}. The task is to verify that all paths satisfy the
following fairness property: if the top floor is infinitely often requested then it is infinitely often
served. This property can be formalised in the modal $\mu$-calculus with alternation depth 2, and the
resulting benchmarking parity games are obtained as $\mu$-calculus model checking games. 

We consider two different implementations of the system, differing in the way that the request list
is stored: in FIFO or in LIFO style. Note that the FIFO implementation satisfies this fairness property
because any requested floor will eventually be served whereas the LIFO implementation does not satisfy
it. This is because new requests will be served before older ones and therefore requests can starve.

With the FIFO system, both local algorithms need to explore the whole game in order to detect that
the universally path quantified formula is satisfied or, equivalently, to find a winning strategy for
player $0$ which comprises the entire game graph. It is not too surprising that global SI with 
all its optimisations outperforms both of them in this case. But note that on the LIFO system, both 
local algorithms explore only a tiny part of the whole game and hence outperform the globally operating 
algorithm. Moreover, local SI is even much faster than the local model checker.
See Fig.~\ref{fig:mucalculusmodelchecking} for all the runtime results.

\paragraph*{LTL Model Checking.} 
We take the well-known dining philosophers problem as a system to be verified. It consists of $n$
forks, each of which can be in three different states (on the table or held by the left or right
philosopher). Transitions are being done by lifting up or placing down one fork. Philosopher $i$ eats
as soon as he holds two forks. We check the LTL formula $\bigvee_{i=1}^n
\mathtt{FG} \mathit{eats}_i$ on the system which asks for a scheduling of the fork lifting that
enables one of them to eventually eat all the time. Note that LTL model checking can also be
viewed as parity game solving \cite{LS-JLC2002}.

Here the local algorithms only need to find a particular scheduling out of all possible ones which enables 
one of the philosophers to eventually eat all the time. Both the local SI and the local model checker 
outperform the global algorithm, again just because they only explore a small part of the whole game graph.
Furthermore, local SI is remarkably more efficient than the local model checker in these test cases,
i.e.\ it explores only a fractional part of the entire game in comparison and therefore needs much
less time. See Fig.~\ref{fig:ltlmodelchecking} for all the runtime results.

\begin{figure}[h]
\centering
{\scriptsize
\begin{tabular}[ht]{|l|r|r||r|r|r||r|r||r|r|}
 \hline
 \multicolumn{3}{|c||}{} & \multicolumn{3}{c||}{\rule[-2mm]{0pt}{6mm} global SI} & \multicolumn{2}{c||}{ local MC} & \multicolumn{2}{c|}{ local SI} \\
  & $n$ & \multicolumn{1}{c||}{$|\mathrm{Game}|$} & \multicolumn{1}{c|}{$t_\mathrm{generate}$} & 
  \multicolumn{1}{c|}{$t_\mathrm{iterate}$} & \multicolumn{1}{c||}{$t_\mathrm{solve}$} & \multicolumn{1}{c|}{visited} & \multicolumn{1}{c||}{$t_\mathrm{solve}$} & \multicolumn{1}{c|}{visited} & \multicolumn{1}{c|}{$t_\mathrm{solve}$} \\
 \hline\hline
 & $9$ & $1,084,808$ & $29.48$s & $28.19$s & $57.67$s & $33,599$ & $0.95$s & $3,439$ & $0.75$s \\
 & $10$ & $3,615,173$ & $177.93$s & $329.53$s & $507.46$s & $90,101$ & $3.05$s & $4,374$ & $1.21$s \\
 \enspace Philo- & $11$ & $11,927,966$ & $1427.76$s & $2794.55$s & $4222.31$s & $201,566$ & $8.16$s & $5,450$ & $2.04$s \\
 \enspace sophers & $12$ & $?$ & $\dagger$ & $-$ & $-$ & $854,922$ & $70.38$s & $6,673$ & $4.32$s \\
 & $13$ & $?$ & $\dagger$ & $-$ & $-$ & $2,763,751$ & $525.45$s & $8,052$ & $6.82$s \\
 & $14$ & $?$ & $\dagger$ & $-$ & $-$ & $-$ & $\dagger$ & $9,595$ & $10.00$s \\
 \hline
\end{tabular}}
\caption{LTL Model Checking}
\label{fig:ltlmodelchecking}
\end{figure}

\paragraph*{Random Games.}
Finally, we consider games generated by a random model distributed with \textsc{PGSolver}. It
essentially creates small random games from a uniform distribution and connects them randomly in order
to obtain a non-trivial SCC structure and therefore to prevent the creation of untypically special games
\cite{fl-atva09}. The initial node is the --- arbitrarily chosen --- first node in the games representation.
The results in Fig.~\ref{fig:randomgames} show that local SI is again much better
than global SI and the local model checker on these games. Note that the number of visited nodes given
there --- as well as the running times --- is the average over 100 runs and therefore not necessarily
an integer value. Also, these random games cannot be created locally. Thus, the times given for all
algorithms are the pure solving times without generation. 

\begin{figure}[h]
\centering
{\scriptsize
\begin{tabular}[ht]{|lr|r||r||r|r||r|r|}
 \hline
 \multicolumn{3}{|c||}{} & \multicolumn{1}{c||}{\rule[-2mm]{0pt}{6mm} global SI} & \multicolumn{2}{c||}{ local MC} & \multicolumn{2}{c|}{ local SI} \\
  &  & \multicolumn{1}{c||}{$|\mathrm{Game}|$} & \multicolumn{1}{c||}{$t_\mathrm{solve}$} & \multicolumn{1}{c|}{visited} & \multicolumn{1}{c||}{$t_\mathrm{solve}$} & \multicolumn{1}{c|}{visited} & \multicolumn{1}{c|}{$t_\mathrm{solve}$} \\
 \hline\hline
 \multicolumn{2}{|l|}{} & $1,000$ & {$0.03$s} & $37.93$ & $0.00$s & $93.02$ & $0.01$s \\
 \multicolumn{2}{|l|}{} & $2,000$ & {$0.07$s} & $103.99$ & $0.01$s & $211.05$ & $0.02$s \\
 \multicolumn{2}{|l|}{} & $5,000$ & {$0.24$s} & $159.14$ & $0.11$s & $255.78$ & $0.03$s \\
 \multicolumn{2}{|l|}{} & $10,000$ & {$0.33$s} & $592.42$ & $0.95$s & $430.36$ & $0.08$s \\
 \multicolumn{2}{|l|}{\enspace Random} & $20,000$ & {$0.77$s} & $6794.45$ & $8.73$s & $453.05$ & $0.10$s \\
 \multicolumn{2}{|l|}{\enspace games} & $50,000$ & {$2.98$s} & $-$ & $\dagger$ & $669.90$ & $0.26$s \\
 \multicolumn{2}{|l|}{} & $100,000$ & {$6.53$s} & $-$ & $\dagger$ & $1001.80$ & $0.48$s \\
 \multicolumn{2}{|l|}{} & $200,000$ & {$17.92$s} & $-$ & $\dagger$ & $1154.40$ & $1.15$s \\
 \multicolumn{2}{|l|}{} & $500,000$ & {$68.33$s} & $-$ & $\dagger$ & $3443.40$ & $5.82$s \\
 \hline
\end{tabular}}
\caption{Random Games}
\label{fig:randomgames}
\end{figure}

\section{Further Work}
\label{sec:concl}

We have presented an on-the-fly strategy improvement algorithm for solving parity games and
shown that it can outperform existing global and even other local methods by several orders
of magnitude. To be precise, the local SI algorithm presented here beats global algorithms in cases
where a winning strategy does not stretch over the entire game graph, even when the global
algorithms are embedded into the optimising framework of the generic engine in \textsc{PGSolver}.

It remains to see which of such optimisations which have been shown to aid global solving 
significantly \cite{fl-atva09} can be used in order to tune local SI, and whether these will also 
act as optimisations in this context rather than a slow-down. For instance, it may be possible to 
do the SCC decomposition of a game graph in a top-down fashion. 

Note that in the expansion step, the local SI algorithm -- as it is presented here -- picks some
node with a yet unexplored subgraph behind it and continues the exploration with this node. It is
of course possibly to parametrise the algorithm with respect to the choice of this node. To be
precise, one could consider yet expansion policies which -- like the improvement policies -- guide
the algorithms choice of such nodes. However, note that the only information that these policies
would have to rely on in order to choose a good node is a current node's successor names, priorities
and their evaluation. The question whether or not one of them is good for the choosing player, however,
depends on the structure of the subgraph lying behind this node rather than on this information. 
Naturally, this subgraph is not visible to the policy for otherwise the algorithm would not be
local. We therefore suggest to simply pick such a successor by random choice.
 
A natural question that also arises is how to localise other global parity game solvers,
in particular Zielonka's recursive one \cite{TCS::Zielonka1998} and Jurdzi{\'n}ski's 
small progress measures algorithm \cite{Jurdzinski/00}. We currently see no method of doing this
in a way that could improve over the global variants like local SI improves over global SI.


\bibliographystyle{eptcs}
\bibliography{./literature}

\end{document}